\def\year{2018}\relax
\documentclass[letterpaper]{article} 
\usepackage{aaai18}  
\usepackage{times}  
\usepackage{helvet}  
\usepackage{courier}  
\usepackage{url}  
\usepackage{graphicx}  
\usepackage{subfigure}
\usepackage[noend]{algpseudocode}
\usepackage{tikz}
\usepackage{multirow}
\usepackage{flushend}
\usepackage{amsmath}
\usepackage{amssymb}
\usepackage{amsfonts}
\usepackage{algorithm}
\begin{document}
%
\title{Measuring the Popularity of Job Skills \\ in Recruitment Market: A Multi-Criteria Approach}
\author{Tong Xu$^{1,2}$, Hengshu Zhu$^{2,}$\thanks{Corresponding Author.}, Chen Zhu$^2$, Pan Li$^{1,2}$, and Hui Xiong$^{1,3,*}$\\
$^1$Anhui Province Key Lab of Big Data Analysis and Application, University of Science and Technology of China\\
tongxu@ustc.edu.cn, lp970625@mail.ustc.edu.cn\\
$^2$ Baidu Talent Intelligence Center, Baidu Inc.\\
\{zhuhengshu, zhuchen02\}@baidu.com\\
$^3$ Rutgers Business School, Rutgers University\\
hxiong@rutgers.edu
}

\maketitle
\begin{abstract}
To cope with the accelerating pace of technological changes, talents are urged to add and refresh their skills for staying in active and gainful employment. This raises a natural question: what are the right skills to learn? Indeed, it is a nontrivial task to measure the popularity of job skills due to the diversified criteria of jobs and the complicated connections within job skills. To that end, in this paper, we propose a data driven approach for modeling the popularity of job skills based on the analysis of large-scale recruitment data. Specifically, we first build a job skill network by exploring a large corpus of job postings. Then, we develop a novel \textbf{S}kill \textbf{P}opularity based \textbf{T}opic \textbf{M}odel (\textbf{SPTM}) for modeling the generation of the skill network. In particular, SPTM can integrate different criteria of jobs (e.g., salary levels, company size) as well as the latent connections within skills, thus we can effectively rank the job skills based on their multi-faceted popularity. Extensive experiments on real-world recruitment data validate the effectiveness of SPTM for measuring the popularity of job skills, and also reveal some interesting rules, such as the popular job skills which lead to high-paid employment.

\end{abstract}

\section{Introduction}
With the coming of knowledge economy era, the competition for skilled talents becomes extremely severe. Indeed, talents who are qualified with employer-targeted skills will get hired easily and obtain attractive reward, such as high compensation and quick promotion. However, there still exists a ``Skill Gap''~\cite{skillgap} between employers and job seekers. Specifically, on the one hand, companies are urgent to seek highly-skilled talents, especially when their demands of job skills change frequently due to the severe business campaigns. On the other hand, to compete against other candidates and outstand from the large personnel pool, job seekers have to acquire certain job skills that target companies pay the most attention on. Such situation is especially significant in massive-population countries like China, where 6 million fresh-graduate students enter the job market every year~\cite{chinagraduate}, far exceeding the number of available jobs on the recruitment market. Therefore, it is very appealing to timely measure the popularity of job skills in recruitment market to guide the employment.

Indeed, the analysis of job skills with measuring their popularity is not novel. Currently, many organizations like Linkedin~\cite{linkedin} and ComputerWorld~\cite{computerworld} have released annual reports for the hottest job skills, which provide insightful guidance for job seekers. However, there still exists several critical challenges along this line. First, in practice, different people tend to have different criteria for their dream jobs, e.g., some applicants pursue high salary, so they choose the big companies in huge cities, while some others tend to select small companies in hometown, even with lower income. Second, job skills are not isolated, but mutually connected to form the latent connections, e.g., skills like ``Javascript'', ``PHP'' and ``Node.js'' usually appeared together, as they are all necessary for the ``Web Front-End Development''. Therefore, hierarchical structure with mutual connections should be integrated when measuring popularity of jobs skills.

To address the aforementioned issues, in this paper, we propose a data-driven approach for measuring the popularity of the job skills based on the analysis of large-scale recruitment data. Specifically, we first build a job skill network ``\textbf{Skill-Net}'' by exploring a large corpus of job postings in recruitment market. Then, we develop a novel \textbf{S}kill \textbf{P}opularity based \textbf{T}opic \textbf{M}odel (\textbf{SPTM}) for modeling the generation of job skills. Particularly, SPTM can integrate different criteria of jobs (e.g., salary level, company size), skill categories, and the latent connections within skills. Therefore, with the help of SPTM, we can effectively rank the job skills based on their multi-perspective popularity. Finally, we evaluate the proposed model with real-world recruitment data and extensive experiments. Specifically, the contribution of this paper can be summarized as follows:

\begin{itemize}
  \item We explore to measure the popularity of job skills in recruitment market with a multi-criteria perspective, which can help to eliminate the ``Skill Gap'' between employers and job seekers in talent recruitment.
  \item We propose a novel SPTM model based on the analysis of large-scale recruitment data, which can effectively model the relations between categorical skills and job criteria.
  \item We evaluate the proposed model with real-world recruitment data and extensive experiments. The experimental results clearly validate the effectiveness of SPTM, and further reveal some interesting discoveries.
\end{itemize}

\begin{table*}[!t]
\centering
\footnotesize
\caption{A toy example of our job posting dataset.}
\begin{tabular}{ccccccc}
\hline
Post ID & Company Scale & Salary & Location & Financial Round & Work Type & Job Description \\ \hline
1000037 & Medium & Low & Huge Cities & A Round & Fulltime & Familiar with Python... \\
1000046 & Medium & Very Low & Huge Cities & B Round & Fulltime & Development of Android...\\
1000082 & Small & Low & Big Cities & D Round & Fulltime & CS Bachelor Degree required...\\
1000144 & Very Big & High & Huge Cities & Listed & Fulltime & Web-design oriented...\\
1000268 & Big & Medium & Normal Cities & D Round & Fulltime & Sufficient with Java...\\
1000462 & Big & High & Huge Cities & Listed & Fulltime & Programming within Linux...\\
\hline
\end{tabular}
\label{Job Posting Dataset}
\end{table*}

\begin{figure}[!b]
\centering
\subfigure[Location]{
\includegraphics[width=3.8cm]{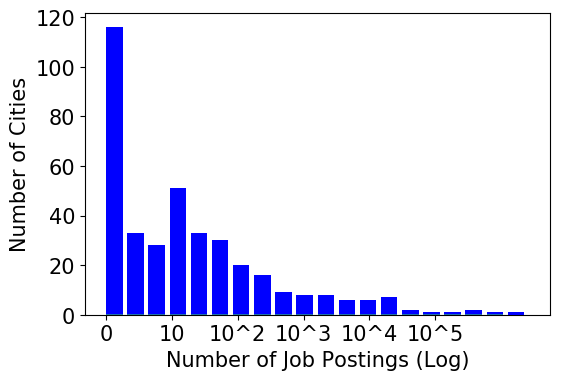}}\hspace{-2mm}
\subfigure[Company]{
\includegraphics[width=4cm]{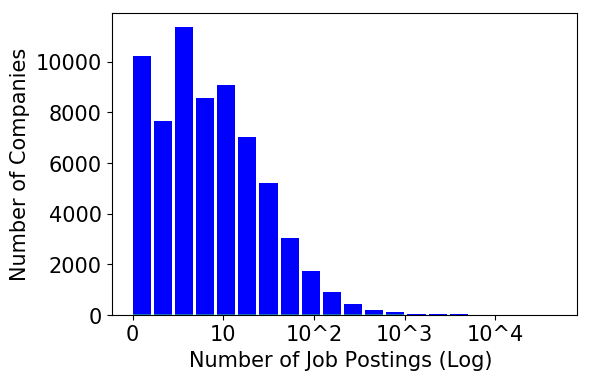}}
\subfigure[Salary]{
\includegraphics[width=3.8cm]{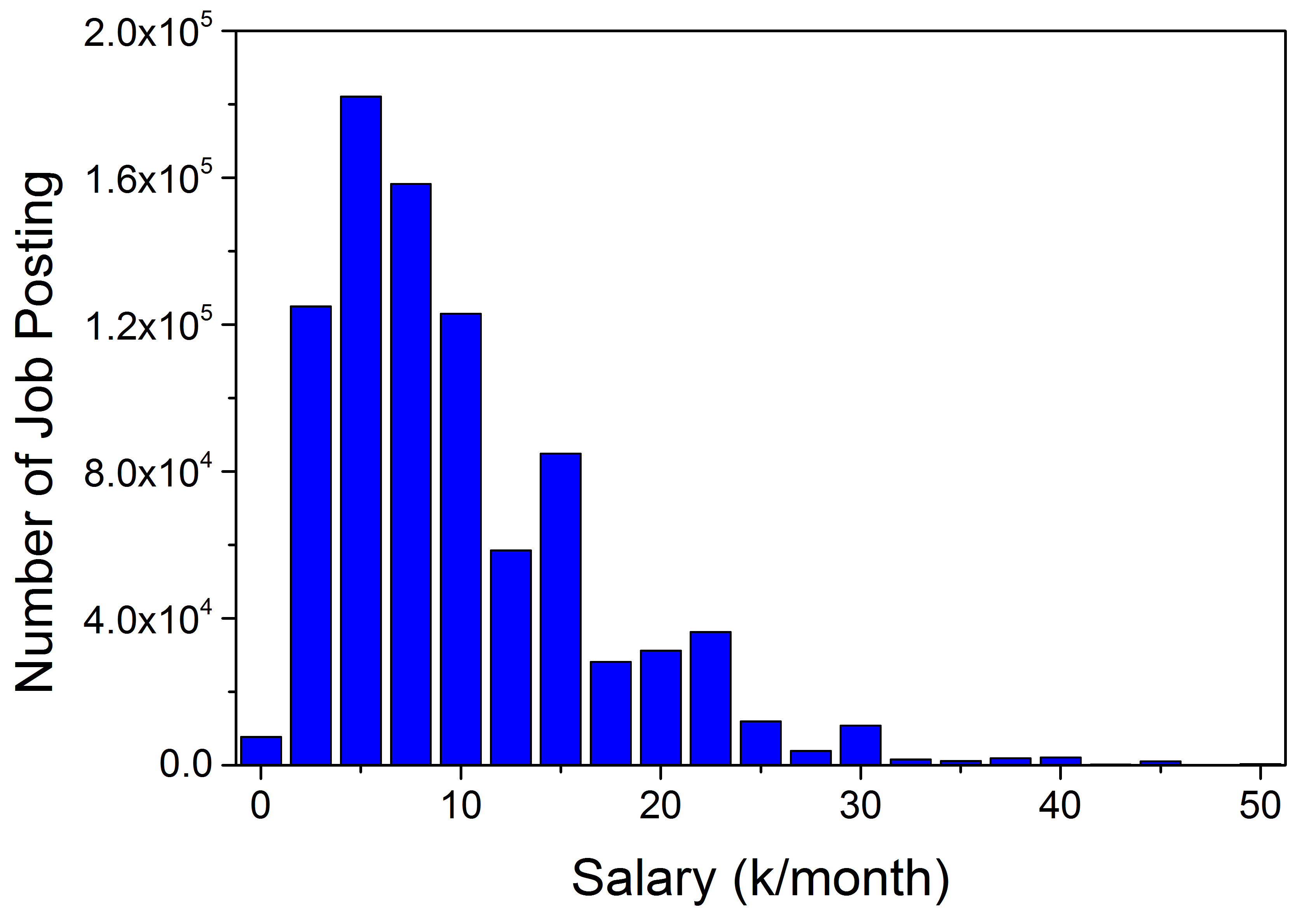}}\hspace{-2mm}
\subfigure[Post Date]{
\includegraphics[width=4cm]{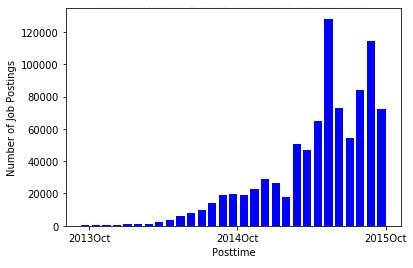}}
\caption{The distributions of job postings with respect to different information categories.}
\label{histgram}
\end{figure}

\section{Data Description}
In this paper, we attempt to develop a data-driven approach for comprehensively measuring the job skill popularity. Specially, to deal with this task, we collected two real-world recruitment data sets, including a large-scale job posting data set with detailed requirement for providing job criteria and training model, as well as a pre-defined categorical skill list to generate the ``Skill-Net".

\subsection{Job posting Dataset}
\label{sec:JD}
To be specific, we extracted the job posting dataset from a famous online recruiting market~\cite{Zhu2016Recruitment}, ranging from 2013 to 2015, which contains 892,454 job postings in total. Among them, 381,578 records are selected with including at least one detailed skill requirement in their job descriptions. Some examples of job posting in our dataset is presented in Table \ref{Job Posting Dataset}, in which we mainly list the criteria of jobs, while the details of job requirement is omitted. Besides, the distributions of job postings with respect to different information categories are illustrated in Figure~\ref{histgram}.

Specifically, to summarize the criteria of jobs, we conclude 5 categories of criteria labels as follows:

\begin{itemize}
\item \textbf{Salary}: Divided into 5 levels (Very High, High, Medium, Low, Very Low), corresponding to monthly salary $>$30k, 20k-30k, 10k-20k, 5k-10k, $<$ 5k respectively.
\item \textbf{Company Scale}: Divided into 5 categories (Very Big, Big, Medium, Small, Very Small), corresponding to the number of employees $>$2000, 500-2000, 100-500, 50-100, $<$ 50, respectively.
\item \textbf{Location}: Divided into 3 categories (Huge Cities, Big Cities, Normal Cities), corresponding to the scale of located cities.
\item \textbf{Financing Round}: Divided into 7 categories (Angel, A, B, C, D, Listed, Unknown), corresponding to the different status of financial round.
\item \textbf{Work Type}: Divided into 3 categories (Fulltime, Part-time, Intern), corresponding to the type of jobs.
\end{itemize}

\subsection{Skill List and ``Skill-Net"}
\begin{figure}[t]
\centering
\subfigure[Frequency of skills]{
\includegraphics[width=4cm]{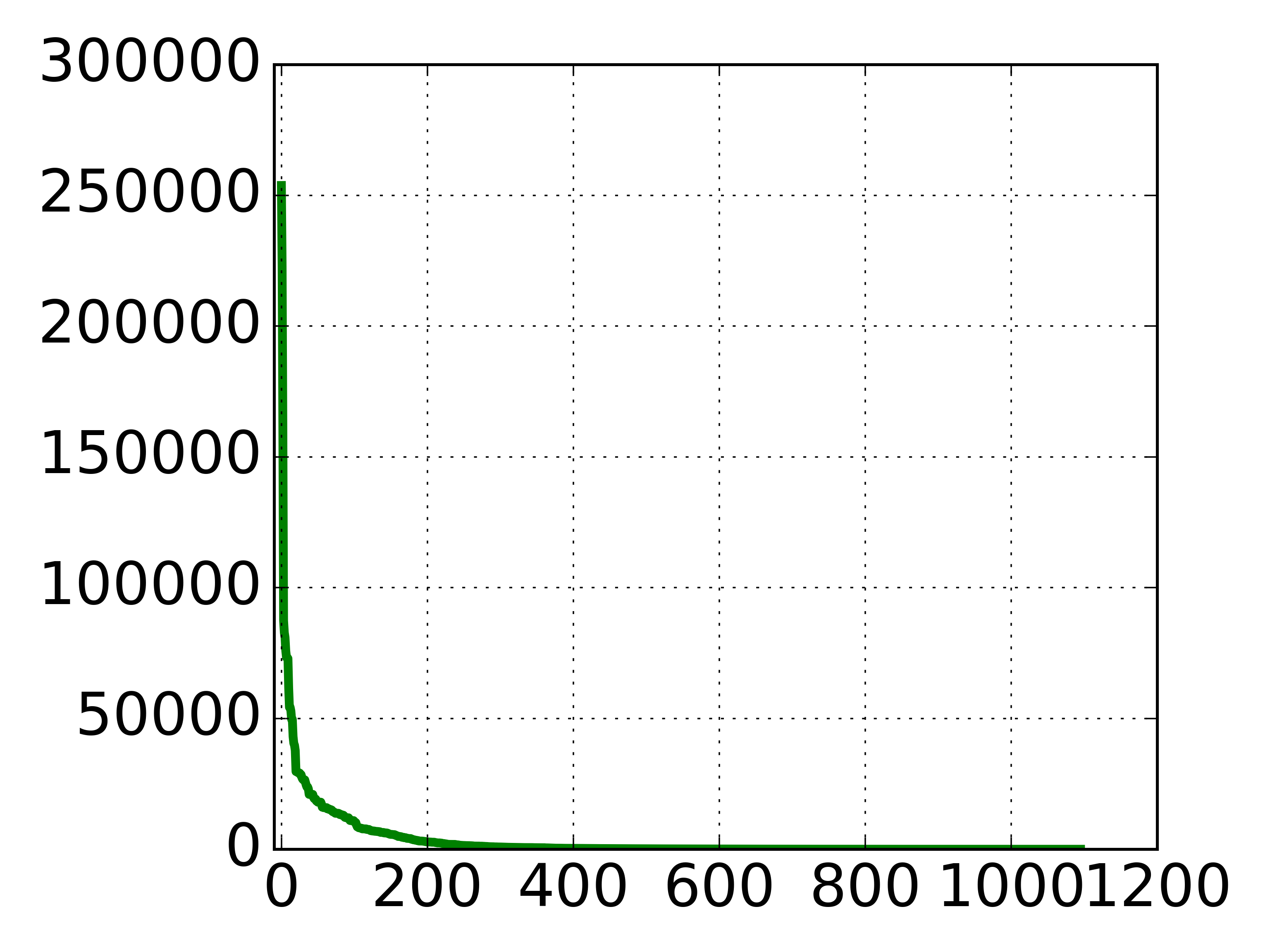}}\hspace{-2mm}
\subfigure[Frequency of skills - Log]{
\includegraphics[width=4cm]{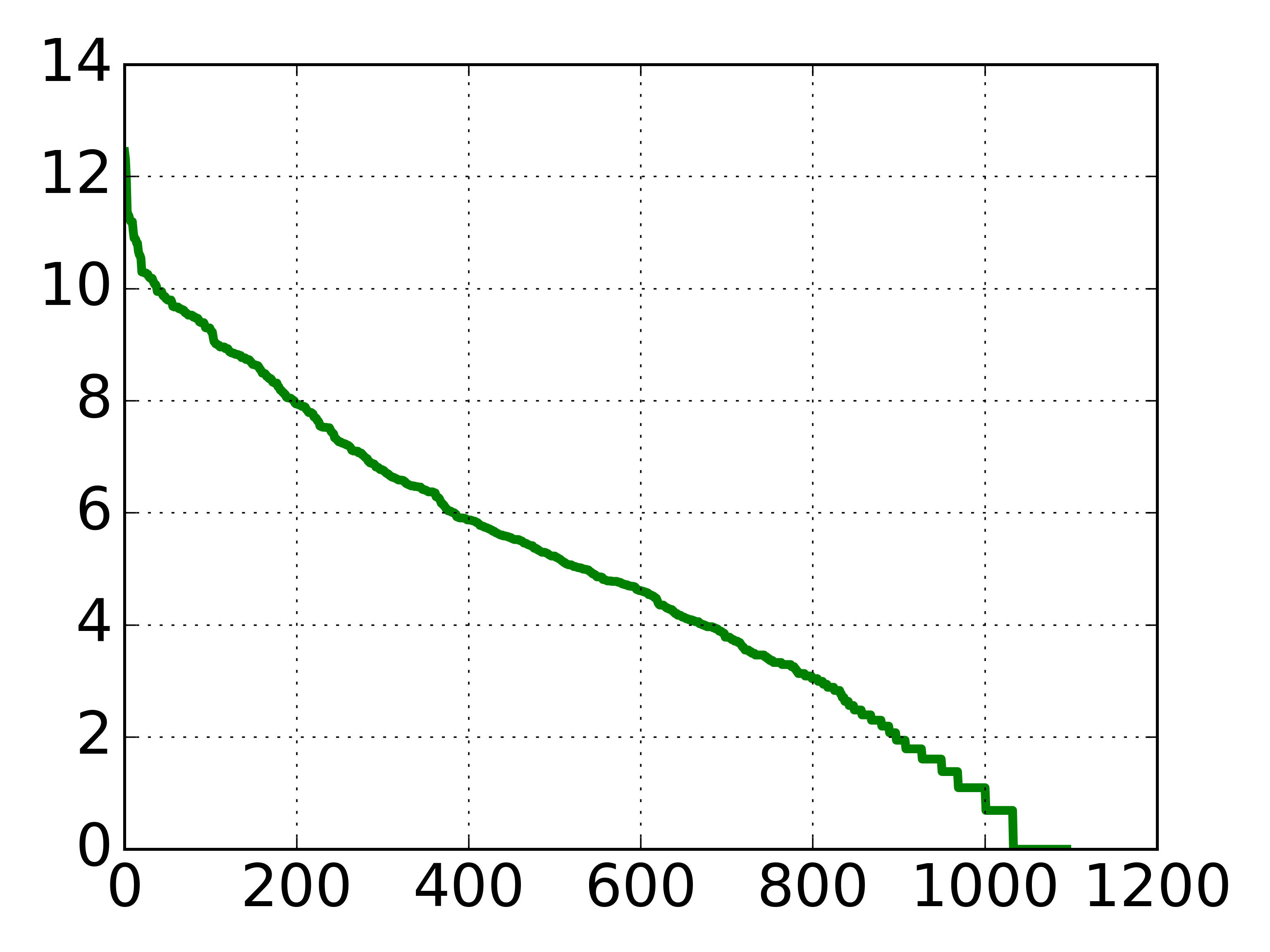}}\hspace{-2mm}
\caption{The distribution of skills frequency, in which Y-axis indicates how many postings contain the skill.}
\label{skill distribution}
\end{figure}

At the same time, for describing the job skills, we extracted a list of pre-defined skills from an online IT community in China~\cite{skillbase}, including two levels of skills, i.e., the \textbf{skill categories}, and the \textbf{detailed skills} which belongs to one category. Totally, 1,729 skills within 54 categories were extracted. To ensure the quality of our analysis, we have removed several detailed skills that are not well defined or too sparse to be shown in a job description, and then built the standard ``\emph{dictionary}" of job skills for modeling. The frequencies of skills in our job posting dataset and its log form ($e$ as the base) are shown in Figure~\ref{skill distribution}, which indicates the skills suffer a significant \emph{long tail effect}, i.e., only a few job skills are popular in recruitment market.

\begin{figure}[t]
\centering
\includegraphics[width=6cm]{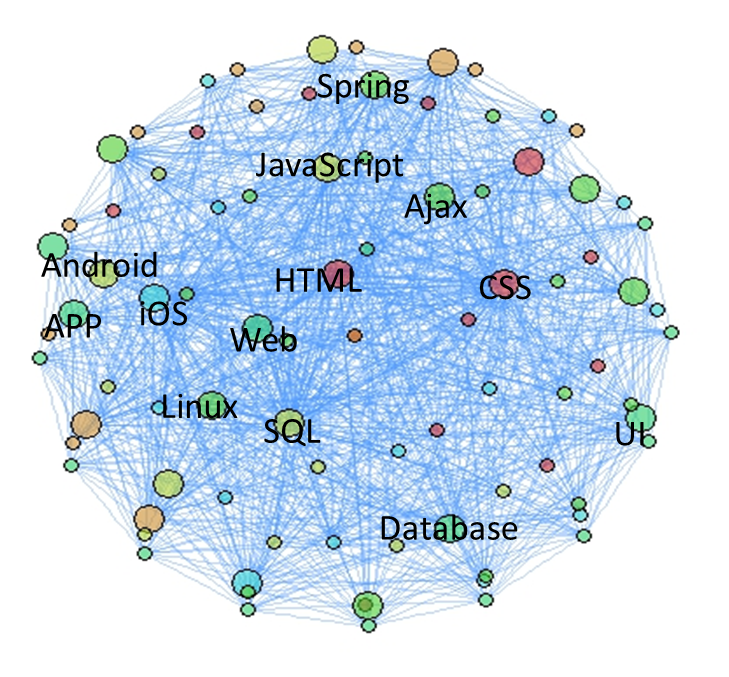}
\caption{The snapshot of ``Skill-Net''.}
\label{Skill Net}
\end{figure}

At the same time, this data set will not only provide the category of skills, but also help to generate the ``\textbf{Skill-Net}''. As mentioned above, job skills are not isolated, but mutually connected following latent hierarchical structure, which cannot be neglected when measuring skill popularity. Unfortunately, no exposed relationship between skills could be directly extracted from original data. Thus, in this paper, based on the standard ``dictionary" above, we pick up the pre-defined skills in each job posting, and then build the ``Skill-Net'' following the heuristic method which is commonly used in social network analysis~\cite{Xu-KDD-2016}. Particularly, we assume that if two skills co-occur in the same job posting, they will be mutually connected. Figure~\ref{Skill Net} shows a snapshot of the ``Skill-Net'', which is generated based on our real-world recruitment dataset.

\section{Popularity Modeling for Job Skills}
\begin{figure*}[!t]
\centering
\includegraphics[width=17cm]{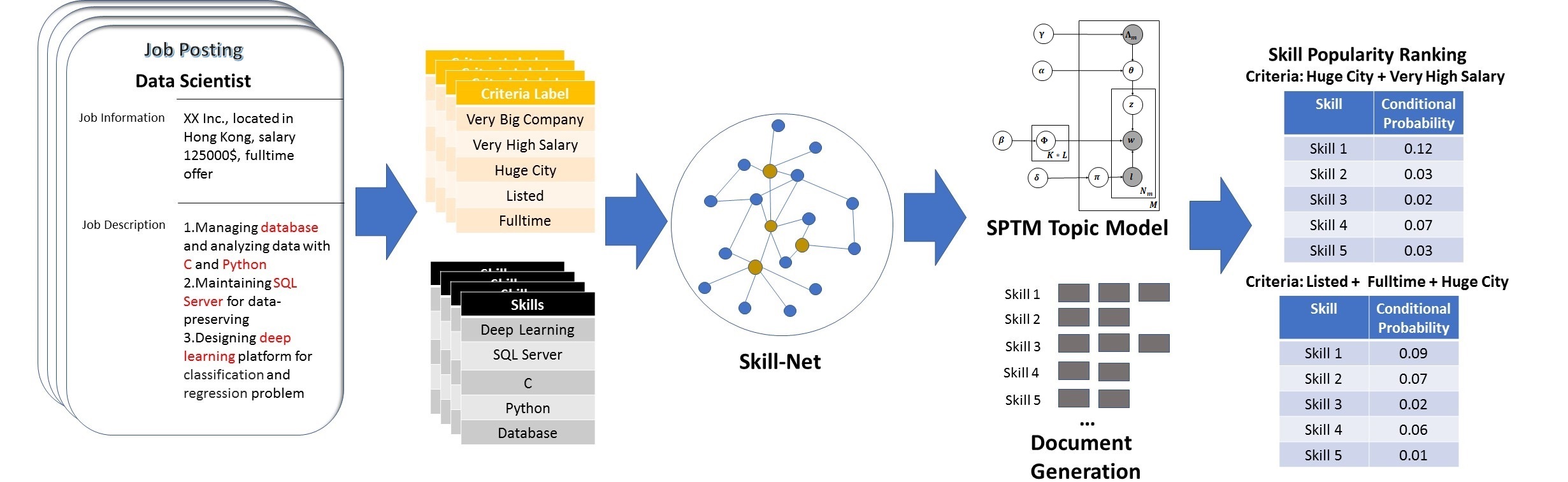}\\
\caption{The overview of our approach for measuring the popularity of job skills.}
\label{jd}
\end{figure*}

As the data sets and ``Skill-Net'' are prepared, in this section, we turn to introduce the details of our novel SPTM approach for effectively measuring the popularity of job skills. To facilitate the understanding, we illustrate the overview of our approach for measuring the skill popularity in Figure~\ref{jd}.

\begin{figure}[t]
\centering
\includegraphics[width=6.5cm]{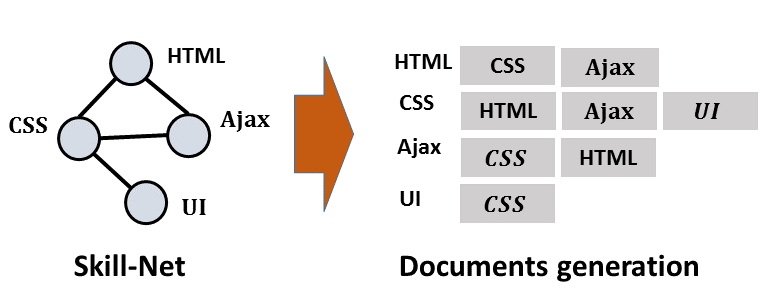}
\caption{The generation of documents in SPTM.}
\label{generation}
\end{figure}

\subsection{Overview of SPTM Model}
Generally, we target at measuring the popularity of skills under different criteria (e.g., salary level, company size, etc.). Therefore, we propose the SPTM approach to learn the latent skill topics combining multi-criteria and hierarchical categories of job skills, which could be used to estimate the popularity of a job skill under certain criteria, via estimating the probability of skill in corresponding topic.

Specially, in SPTM, we treat each skill, named \textbf{central skill} in our paper, as a ``\emph{document}'' in ``Skill-Net''. Also, those neighboring skills, named \textbf{skill tokens}, are regarded as corresponding ``\emph{word tokens}'' within, which is similar with the basic idea of Topic Model. Figure~\ref{generation} shows an example of the document generation process in SPTM, in which each line of the right part indicates a ``document''. Besides, as we known each job can be classified by some standard criteria. And we further think if a central skill has appeared in a job posting, it should has some relationships with the criteria labels of the job (e.g., very high salary, very big company, etc.). By this assumption, we generate a set of \textbf{criteria labels} for each central skill.

Just as in LDA, we also use $\theta_{m}$ and $\phi_{k}$ to represent the distribution over topics for $m$-th central skill and the distribution over skills for $k$-th topics.
Besides, to model the connection between criteria labels and central skills, we follow the ideas in~\cite{ramage2009labeled} and associate each topic with a criteria label. Specially, we have $\Lambda_{m_{\star}}$ to presents the criteria labels of $m$-th central skill, where $\Lambda_{m_{\star}}$ is a binary topic presence/absence vector and each $\Lambda_{m_{k}} \in \{0,1\}$. Then to keep the above hypothesis, we generate the $m$-th central skill's distribution over topic $\theta_{m}$ by $Dir(\cdot|\alpha \times \Lambda_{m_{\star}})$.

Besides, according to our basic assumptions, the popularity of skills will be affected not only by mutual connections and criteria, but also corresponding skill categories, which can further help us understand hierarchical skill relationship from macro perspective. We follow the ideas in~\cite{huai2014toward} and hypothesize each central skill also has a distribution over skill categories. $l_{s}$ is used to represent the skill category of skill $s$ (e.g., the category of skill ``CNN'' is ``machine learning''). Thus the skill token $w_{m,i}$ is finally generated by $Multi(\cdot|\phi_{z_{m,i},l_{m,i}})$.

The detailed generation process of SPTM is illustrated in Algorithm~\ref{alg:SPTM}, and related mathematical notations are summarized in Table~\ref{notation}.

\begin{table}[t!]
\centering
\caption{Mathematical Notations.}\label{notation}
\footnotesize
\begin{tabular}{cc}
\hline
\textbf{Symbol} & \textbf{Description} \\ \hline
$\alpha$ & Dirichlet prior for generating $\theta$ \\
$\beta$ & Dirichlet prior for generating $\phi$ \\
$\delta$ & Dirichlet prior for generating $\pi$ \\
$\gamma$ & Distribution parameters for generating $\Lambda_{m_{\star}}$ \\
$\pi$ & Distribution parameters for generating $\Lambda_{l_{\star}}$ \\
$\Lambda_{m_{\star}}$ & Topic presence/absence vector of $m$-th central skill \\
$\theta_{m}$ & Distribution over topics for $m$-th central skill \\
$z_{m,i}$ & Topic label of $i$-th skill token of $m$-th central skill \\
$l_{m,i}$ & skill category of $i$-th skill token of $m$-th central skill \\
$l_{s}$ & skill category of $s$-th skill \\
$w_{m,i}$ & $i$-th skill token of $m$-th central skill \\
$\phi_{k,l}$ & Distribution over skills for $k$-th topic and $l$-th category \\
$K$ & Number of topics/criteria \\
$S$ & Number of skills \\
$L$ & Number of skill categories \\
$M$ & Number of central skills \\
$N_m$ & Number of skill tokens of $m$-th central skill \\
\hline
\end{tabular}
\end{table}

\begin{algorithm}[t]
\centering
\footnotesize
\caption{The Generation Process of SPTM.}\label{alg:SPTM}
\begin{algorithmic}[1]
\For {each topic $k \in \left\{{1,2,\cdots,K}\right\}$}
\State Generate $\phi_{k} \sim Dir(\cdot|\beta)$
\EndFor

\For {each central skill $m \in \left\{{1,2,\cdots,M}\right\}$}
\For {each topic $k \in \left\{{1,2,\cdots,K}\right\}$}
\State Generate $\Lambda_{m_{k}} \in \{0,1\} \sim Bernoulli(\cdot|\gamma_{k})$
\EndFor
\State Generate $\pi_{m} \sim Dir(\cdot|\delta)$
\State Generate $\theta_{m} \sim Dir(\cdot|\alpha \times \Lambda_{m_{\star}})$
\For {each skill token $i \in \left\{{1,2,\cdots,N_{m}}\right\}$}
\State Generate $z_{m,i} \sim Multi(\cdot|\theta_{m})$
\State Generate $l_{m,i} \sim Multi(\cdot|\pi_{m})$
\State Generate $w_{m,i} \sim Multi(\cdot|\phi_{z_{m,i},l_{m,i}})$
\EndFor
\EndFor
\end{algorithmic}
\end{algorithm}

\subsection{Model Inference}
Then, we will introduce the technical solution of model inference. Specially, we use collapsed Gibbs sampling for training, where the sampling probability of $z_{m,i}$, which indicates the topic of $i$-th skill token in $m$-th central skill, is given by:
\begin{footnotesize}
\begin{eqnarray}
P(z_{m,i}=j|z_{m,-i},w_{m,i},l_{m,i})~~~~~~~~~~~~~~~~~~~~~~~ \nonumber\\
\propto P(w_{m,i},l_{m,i}|z_{m,i}=j,z_{m,-i},w_{m,-i},l_{m,-i})P(z_{m,i}=j|z_{m,-i}),
\end{eqnarray}
\end{footnotesize}

where $w_{m,-i}$ is all of skill tokens of $m$-th central skill except $i$-th one and $l_{m, -i}$ is similar. $z_{m,-i}$ is the topic labels of $w_{m,-i}$.
Here $P(w_{m,i},l_{m,i}|z_{m,i}=j,z_{m,-i},w_{m,-i},l_{m,-i})$ stands for the appearance probability of the skill token $w_{m,i}$ and its category $l_{m,i}$ given $z_{m,i}=j$, and $P(z_{m,i}=j|z_{m,-i})$ represents the probability of $z_{m,i}=j$ in $m$-th central skill regardless of $i$-th skill token. Then we can calculate these two probabilities as follows:
\begin{footnotesize}
\begin{eqnarray}
\footnotesize
&P(w_{m,i},l_{m,i}|z_{m,i}=j,z_{m,-i},w_{m,-i},l_{m,-i}) \nonumber \\
&=\frac{n_{w_{m,i},l_{m,i},j,-(m,i)}+\beta_{w}}{\sum_{s=1}^{S}(n_{s,l_{s},j,-(m,i)}+\beta_{s})} \cdot \frac{n_{m,l_{m,i},-(m,i)}+\delta_{l_{m,i}}}{\sum_{l^{'}=1}^{L}(n_{m,l^{'},-(m,i)}+\delta_{l^{'}})}, \\
&P(z_{m,i}=j|z_{m,-i}) = \frac{(n_{m,j,-(m,i)}+\alpha_{j}) \cdot \Lambda_{m_{j}}}{N_{m}-1+\sum_{k=1}^{K}\alpha_{k}\cdot \Lambda_{m_{k}}},
\end{eqnarray}
\end{footnotesize}
where $n_{s,l,j,-(m,i)}$ is count of $s$-th skill in $l$-th skill category and $j$-th topic regardless of skill token $w_{m,i}$ and $n_{m,j,-(m,i)}$ is count of $j$-th topic appearing in $m$-th central skill expect $w_{m,i}$. $n_{m,l,-(m,i)}$ is the count of $l$-th skill category appearing in $m$-th central skill expect $w_{m,i}$.
After several iterations of Gibbs sampling, we could then estimate the conditional probability of a skill $w$ and the corresponding skill category $l$ within each topic as follows:
\begin{footnotesize}
\begin{eqnarray}
P(w,l|z=j) = P(s|l,z=j)P(l), ~~~~~~~~~~~~~~~\nonumber \\
~~~~~~~~~~~= \frac{n_{w,l,j}+\beta_{w} }{\sum_{s^{'}=1}^{S}(n_{s^{'},l_{s^{'}},j}+\beta_{s^{'}})} \frac{n_{m,l}+\delta_{l}}{\sum_{l^{'}=1}^{L}(n_{m,l^{'}}+\delta_{l^{'}})},
\end{eqnarray}
\end{footnotesize}

After learning and inferencing of SPTM, we could now measure the popularity of a job skill under different criteria. Specially, given a specific group of criteria $\Delta=\{\Lambda_{m_{i}}\}$, in which each indicates a criterion, (e.g. salary, location, company scale, financial round or work type), considering that different criteria are independent with each other, we can estimate the popularity of skill $w$ and the corresponding skill category $l$ via Bayes' Formula by:
\begin{footnotesize}
\begin{eqnarray}
P(w,l|\Delta) = \sum_{\Lambda_{m_i}\in \Delta}\Big(P(\Lambda_{m_i})\cdot ~~~~~~~~~~~~~~~~~~ \nonumber \\
                    ~~~~~~~~~~        \big( \sum_{j=1}^{K}P(z=j|\Lambda_{m_{i}})P(w,l|z=j)  \big)\Big).
\end{eqnarray}
\end{footnotesize}

\section{Experiments}
In this section, we will evaluate the effectiveness of our purposed model. Specially, our SPTM will be validated on real-world recruitment data sets compared with two typical baselines. After that, we discuss about the experimental results with capturing some interesting rules, and a case study will be conducted to further confirm our basic ideas.

\subsection{Experimental Setup}
Firstly, we summarize the details of experimental setup in this subsection, including data pre-processing, parameter setting and baseline methods.

\vspace{2mm}
\noindent\textbf{Data Pre-processing.}
As introduced in Section 2, we conducted our experiments on the real-world dataset collected from an online recruiting market and IT community. To ensure the effectiveness and avoid missing values, we filtered out those job-description records without any detailed skill, which compose of 57.24\% of the original data. Correspondingly, those skills which never appear in job-description records will also be removed.

\vspace{2mm}
\noindent\textbf{Parameter Setting.}
In our SPTM model, three sets of parameters should be preset. First, for the generation process of SPTM, we manually selected the values for Dirichlet prior with ($\delta$=1, $\alpha$,$\beta$,$\gamma$=0.01) based on multiple optimizations. Second, we set to the number of topic $K$ as 23, corresponding to the 23 unique criteria within 5 different categories mentioned in \emph{Data Description}. Finally, for the model training process, the maximal iteration times was set as 800, and threshold of stop condition was set as $10^{-3}$.

\vspace{2mm}
\noindent\textbf{Baseline Methods.}
To validate the performance of SPTM for measuring skill popularity, two typical baseline methods are selected as baseline methods as follows:
\begin{itemize}
\item \textbf{Frequency.} The intuitive technique to measure the popularity of job skills. Specially, we calculated the appearance of each skill under different criteria labels, and then used the normalization of skill frequency as conditional probability for ranking the popularity.
\item \textbf{LLDA.} Labeled-LDA (LLDA)~\cite{ramage2009labeled} is a topic model similar with our SPTM, which however does not consider the skill categories and mutual connections within skills. We used the same feature labels and identical parameters as SPTM for training L-LDA.
\end{itemize}

\subsection{Experimental Results}
For the comprehensive validation of SPTM, two kinds of experiments were executed, i.e., skill-oriented topic evaluation, and job skill recommendation for a given job posting.

\vspace{2mm}
\noindent\textbf{Skill-oriented Topic Evaluation.}
To evaluate the effectiveness of our SPTM approach, especially for the topic modeling performance, we picked up the top 8 skills of each topic listed by all the three techniques, and then asked 5 senior HR experts to judge the correctness. In detail, two factors will be considered, namely the \emph{Validity Measure} (\textbf{VM}) which measures how many topics extracted are \emph{valid} (i.e., at least 4 skills among 8 skills are mutually coherent in job recruitment), and the \emph{Coherence Measure} (\textbf{CM}), which counts how many skills are relevant to the other 7 skills in the same topic. These two metrics could be calculated as follows:
\begin{footnotesize}
\begin{equation}
VM=\frac{\#~of~ \emph{valid}~topics}{\#~of~topics}, CM=\frac{\#~of~\emph{relevant}~skills}{\#~of~skills}.
\end{equation}
\end{footnotesize}

The average results of 5 experts are shown in Table~\ref{Comparison}, in which our SPTM outperforms all the baselines in terms of both VM and CM metrics. Moreover, as we only selected the top 8 skills of each topic, the ranking performance of job skills via our SPTM approach has been also validated.

\vspace{2mm}
\noindent\textbf{Job Skill Recommendation.}
Then, we turn to evaluate the job skill recommendation task. As SPTM could measure the popularity of job skills with different criteria, it is intuitive to recommend appropriate job skills with given job description. Definitely, these results will be beneficial for both employers and job seekers, since they could offer guidance for companies to follow the trends, and also help the candidates to select the proper positions, or prepare for the interview.

Specially, for the evaluation, we select the test set formed by 50,000 job postings excluded from the training set, which contains 623 different combination of criteria, and then use the log-likelihood as measurement to evaluate the performance. The results are shown in Table~\ref{Log Likelihood}, which indicates that our SPTM approach performs better with recommending more appropriate job skills compared with the baselines.

\begin{table}[t]
\centering
\footnotesize
\caption{Average VM/CM Comparison}\label{Comparison}
\begin{tabular}{ccc}
\hline
Model & VM & CM \\ \hline
SPTM & \textbf{0.835} & \textbf{0.625}\\
Labeled-LDA & 0.700 & 0.300 \\
Frequency & 0.575 & 0.325 \\ \hline
\end{tabular}
\label{Comparison}
\end{table}

\begin{table}[t]
\centering
\footnotesize
\caption{Recommendation performance in log-likelihood}
\begin{tabular}{cc}
\hline
Model & log-likelihood \\ \hline
SPTM & \textbf{-6320262.155}\\
Labeled-LDA & -7065396.092\\
Frequency & -6511843.161\\
\hline
\end{tabular}
\label{Log Likelihood}
\end{table}

\subsection{Discussion and Insights}
\begin{figure}[!b]
\centering
\subfigure[Very Big Company]{
\includegraphics[width=4cm]{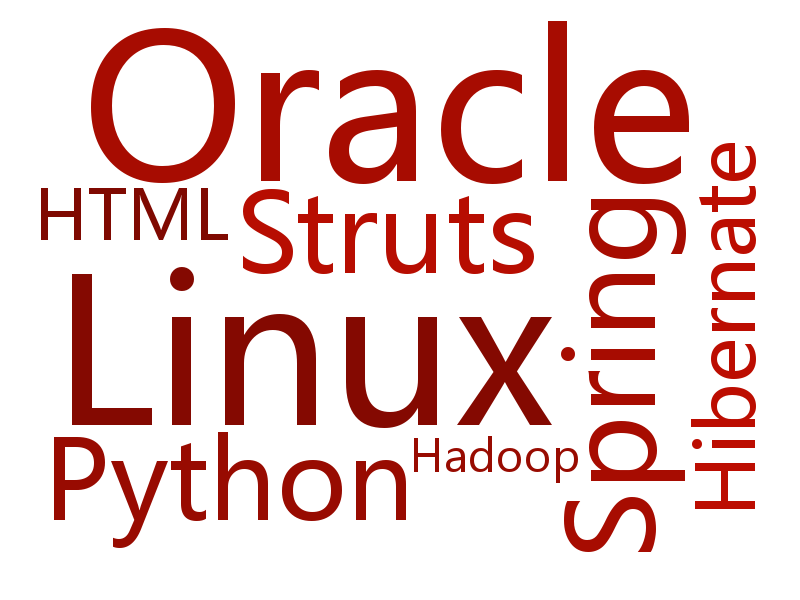}}\hspace{-2mm}
\subfigure[Very Small Company]{
\includegraphics[width=4cm]{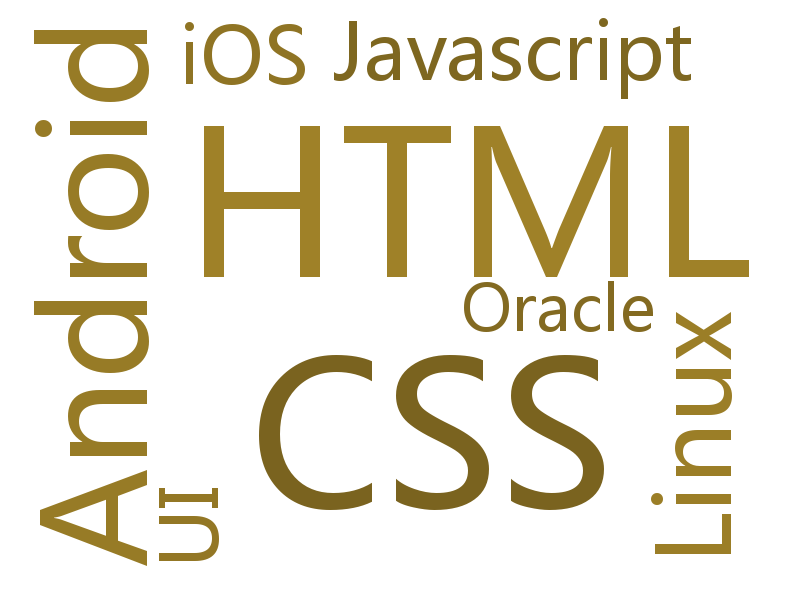}}\hspace{-2mm}
\caption{Top 8 popular job skills (Company Scale).}\label{ResultsComScale}
\end{figure}

\begin{figure}[!b]
\centering
\subfigure[Very High Salary]{
\includegraphics[width=4cm]{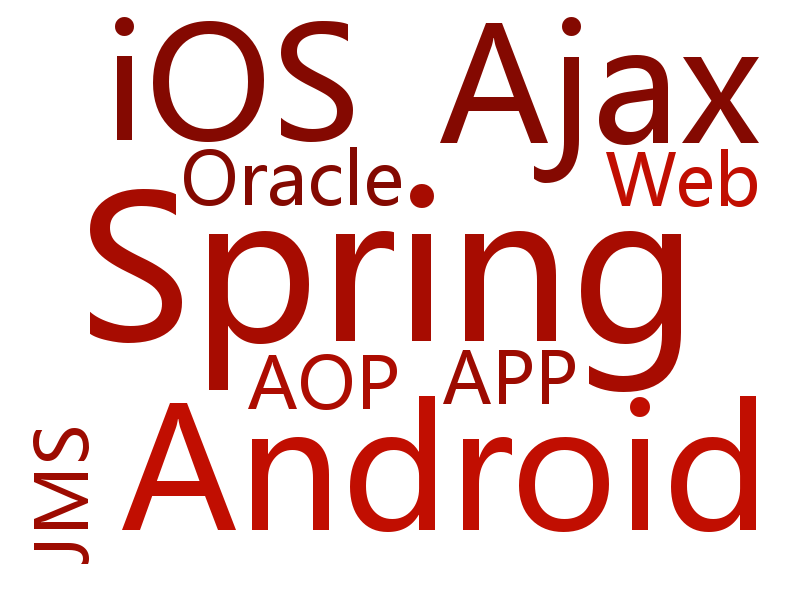}}\hspace{-2mm}
\subfigure[Very Low Salary]{
\includegraphics[width=4cm]{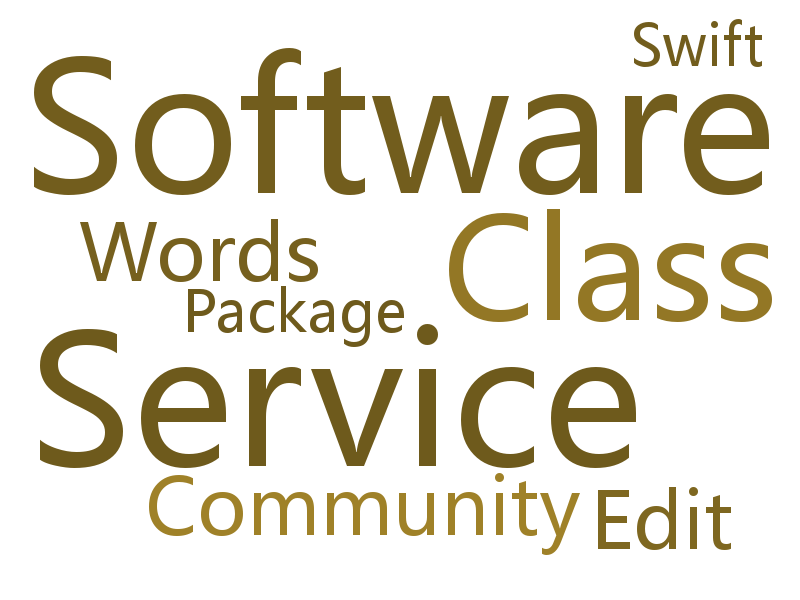}}\hspace{-2mm}
\caption{Top 8 popular job skills (Salary).}\label{ResultsSalary}
\end{figure}

As the effectiveness of SPTM has been validated, in this subsection, we will further discuss about some interesting discoveries as follows. Specially, we list the top 8 skills ranked by our SPTM framwork for different job criteria in Figure~\ref{ResultsComScale} and \ref{ResultsSalary}, in which the size indicates the ranking of skill.

\vspace{2mm}
\noindent $\diamond$ \textbf{Q1: Will different companies lead to different skills?}\\

Indeed, the impact of different companies depends on the business range, which is usually related to the scale of companies. For instance, as shown in Figure~\ref{ResultsComScale}, big companies like Google and Baidu usually hold a wide range of business and large amount of users, which result in the massive data. Thus, technical skills like ``Oracle'' to manage database, ``Python" for data analysis, or even ``Hadoop'' for distributed computing are required. On the contrary, the small companies usually focus on a certain application field, thus ``HTML'' for front-end designing, or ``Android'' and ``iOS'' for mobile App developer are required, which are shown in. Thus, the skills for different company scale could be significantly different.

\vspace{2mm}
\noindent $\diamond$ \textbf{Q2: What are the ``salary-oriented'' skills, and why?}\\

As shown in Figure~\ref{ResultsSalary}, we can find that job skills related to design and development in mobile services, like ``Android'', ``iOS'' and ``Web-App'' tend to receive higher salary, which require not only technical training, but also inspiration and innovation. Meanwhile, the increasing development of mobile internet also stimulate the popularity of corresponding skills. On the contrary, the basic skills with less techniques, like ``Service'' and ``Edit'', lead to relatively lower salary. Clearly, the irreplaceable quality of a certain skill could possibly affect the reward it gains.

\begin{table}[t]
\centering
\footnotesize
\caption{Changes of top 10 popular job skills.}\label{Q4}
\begin{tabular}{c|c|c}
\hline
 2015 & 2014 & 2013 \\ \hline
 SQL & Software & Test \\
 Database & Class & Hardware \\
 IO & Framework & Data Analysis \\
 Framework & Mathematics & Communication \\
 Data Structure & Android & IO \\
 Python & Management & Relation \\
 Search Engine & Data Analysis & Network \\
Software& Algorithm & UI \\
 Network & Data Structure & Framework \\
\hline
\end{tabular}
\end{table}

\vspace{2mm}
\noindent $\diamond$ \textbf{Q3: How does the popularity of job skills change over different years?}\\

In Table~\ref{Q4}, we present the comparison of popular skills in the past three years by jointly considering positive job criteria (i.e., Very High Salary, Very Big Company, Huge City, Listed and Fulltime). Without considering the common skills, like Development and C, which keep relatively stable, we can see a clear increasing demand of data-driven skills compared with several years ago. On the contrary, the network-related skills decline, and we could witness the transfer of market trends from hardware-oriented job skills to software-oriented job skills.

\subsection{Case Study: Will Talents with Popular Skills Get Hired Easily?}
Finally, we will present a case study showing the relationship between our modeling results and realistic recruiting results with a real-world recruitment data set.

\vspace{2mm}
\noindent \textbf{Recruitment Dataset.} The dataset was provided by an IT company in China, which contains 140,757 resumes that have applied for technical jobs in this company. Specially, all the resumes had been scored by the Human Resource Department as follows:
\begin{small}
\begin{itemize}
\item \textbf{Score 0:} \emph{Candidate was failed without interview}.
\item \textbf{Score 1:} \emph{Candidate was failed in the interview}.
\item \textbf{Score 2:} \emph{Candidate was admitted but failed to join the company}.
\item \textbf{Score 3:} \emph{Candidate was admitted and joined the company}.
\end{itemize}
\end{small}

\begin{figure}[b]
\centering
\includegraphics[width=6.5cm]{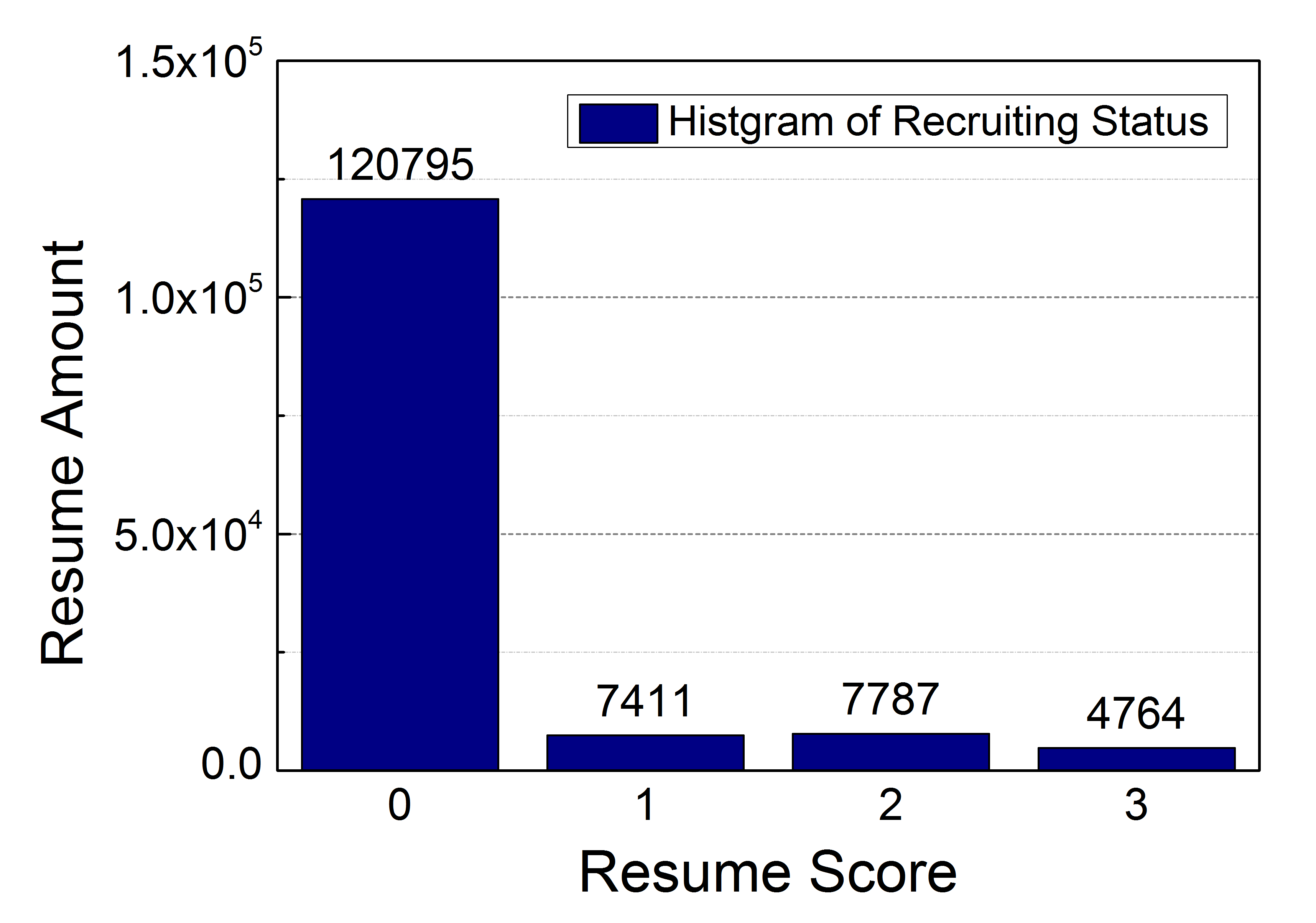}
\caption{The distribution of the number of different resume scores in our recruitment dataset.}
\label{Recruiting Dataset}
\end{figure}
The distribution of each label in our dataset is listed as histogram in Figure~\ref{Recruiting Dataset}, which indicates that around 85\% of candidates are filtered out before receiving interview.

\vspace{2mm}
\noindent \textbf{Correlation Measurement.}
According to our basic assumptions, the candidates with popular job skills will be more probable to be enrolled. As the effectiveness of our model has been confirmed, we now apply SPTM for resume analysis to simulate the recruitment process. To be specific, we first extracted the corresponding skills from resumes, and then calculated the ``skill score'' for each resume, which is defined by the following formula:
\begin{footnotesize}
\begin{align}
Score_{Skill}=\sum_w |w|\times P(w|\Lambda_{m_{\star}}),
\end{align}
\end{footnotesize}
\noindent where $w$ is a job skill appeared in the resume, and $|w|$ is the frequency acting as the weight. Besides, considering the status of this company, the criteria labels $\Lambda_{m_{\star}}$ are set as (Very High Salary, Huge City, Big Compay, Listed).

To validate the performance, we introduce the Spearman correlation coefficient and Kendall's $\tau$ to measure the correlation between ``skill score'' and corresponding resume scores. The correlation measurements are shown in Table~\ref{Correlation Measurement}, in which the ``skill score'' by our SPTM holds much stronger correlation with resume scores than all the baselines.

\vspace{2mm}
\noindent \textbf{Hypothesis Test.}
Besides, hypothesis test on Kendall's $\tau$ indicates that resume ranking obtained by our SPTM is statistically significant, which further proves the practicability of our model. Specially, the sampling distribution of Kendall's $\tau$ could be estimated by the normal distribution, with mean as 0 and variance as $\frac{2(2N+5)}{9N(N-1)}$, where $N$ equals to the number of samples. Thus, we conducted the $z$-test on $\tau$ obtained by SPTM, and the $z$-value could be calculated by the formula below, which validates the result with high confidence:
\begin{footnotesize}
\begin{eqnarray}
z_{value}=\frac{3\times\tau\sqrt{N(N-1)}}{\sqrt{2(2N+5)}} \nonumber \\
=156.748 \ge z_{0.0001}.
\end{eqnarray}
\end{footnotesize}

In summary, based on the above experiments, we prove that talents with popular skills indeed get hired easily. Also, we realize that our SPTM approach has potential for facilitating the process of resume screening in talent recruitment.

\begin{table}[t]
\centering
\footnotesize
\caption{The results of correlation analysis.}
\begin{tabular}{cccc}
\hline
Measurement|Model & SPTM & LLDA & Frequency\\ \hline
Spearman & \textbf{0.3002} & 0.1855 & 0.2015 \\
Kendall's $\tau$ & \textbf{0.2452} & 0.1528 & 0.1603 \\
\hline
\end{tabular}
\label{Correlation Measurement}
\end{table}

\section{Related Work}
In this section, we summarize the related works following three aspects, namely Skill Ranking, Recruitment Market Analysis and Topic Models.

\subsection{Skill Ranking}
Skill ranking of IT techniques has attracted growing interest in the last decade. Previous study in 2005~\cite{Prabhakar2005IT} discussed about the changing tendency of 14 different IT skills using their percentage in job postings during a certain period, where Web programming, Unix, C++, Java, SQL programming, and Oracle database were listed as the top six skills. Recently, some online recruiting companies also publish their annual report for the hottest skills like~\cite{Miller2005Skill} and~\cite{Skomoroch2012SKILL}. For example, in~\cite{linkedin}, Linkedin predicted top skills that can get people hired in 2017, based on the ranking results of hiring and recruiting activities. Also, in~\cite{computerworld}, Computerworld published an article illustrating 10 hottest tech skills for 2017 based on the evaluations and judgements of experts.

However, the prior arts above not only fail to provide description of detailed skills to be learned (instead, they provide only rough category of skills), but also lack of interpretability and scalability in modeling, especially for describing the structure of ``Skill Net''.

\subsection{Recruitment Market Analysis}
Recruitment market analysis has always been an appealing topic for applied business researchers, dating back from Adam Smith and his \emph{The Wealth}, for labor distribution is a crucial element in the development of modern society. Economists look into the problem from two different prospective: Macro prospective~\cite{Encyclopedias2015Encyclopedia}, where Solow proposed his growth model, and others study topics about the demographic structure and participation rate of labor, the relation between inflation and unemployment rate and how labor contributes in gross productivity or expenditures, etc. Micro prospective, where studies are all based on the basic market cleaning framework, stating that all employees choose their best balance between leisure and work, while all employers hire with budget constrain, and consequently the wage is derived as the marginal labor cost.

Recently, researchers are devoted to combining data mining techniques with recruitment market analysis, including offer categorization~\cite{Malherbe2015Bringing}, talent career path analysis~\cite{li2017prospecting}, market trend analysis~\cite{Zhu2016Recruitment,Lin-AAAI-2017}, and talent circles~\cite{Xuhuang-KDD2016}. However, few of them studied the problem of measuring the popularity of job skill in recruitment market, not to mention the multi-faceted popularity ranking, which is the focus of this paper.
.
\subsection{Topic Models}
Topic models have been successfully applied to problems in a variety of fields, such as marketing campaign~\cite{liu2014discovering}, emotion recognition~\cite{zhu2016tracking}, biometrics~\cite{Wang2011Finding}, genetics~\cite{Blei2007Correction}, social media~\cite{xu2014learning} and mobile data mining~\cite{Farrahi2012Extracting,zhu2015popularity}.

In the past years, a number of variations of the classical Latent Dirichlet Allocation (LDA)~\cite{blei2003latent} have been proposed for solving different kinds of problems. To utilize the existing document labels, supervised LDA models~\cite{mcauliffe2008supervised} have shown the effectiveness on classification and regression tasks, while they are limited to assign one topic for each document. To this end, Labeled-LDA~\cite{ramage2009labeled} has emancipated this limitation by allowing multiple topics for a single document, thus outperforms the previous supervised models. However, the above models cannot take the hierarchical information of words into account. Therefore, some hierarchical topic models have been proposed, such as  Hierarchical LDA~\cite{griffiths2004hierarchical}, and LDAC (Latent Dirichlet Allocation on Context)~\cite{Bao2010An}.

In this paper, different from the above studies, the proposed model SPTM can integrate both job criteria labels and skill categories for modeling the generation of job skills, and then measure the popularity. Further, the ``documents" here are generated by ``central skills" and their neighboring nodes in ``Skill-Net'', which reflects the connections within skills.

\section{Conclusion}
In this paper, we proposed a data-driven approach for measuring the popularity of job skills based on the analysis of large-scale recruitment data. Specifically, we first built a job skill network by exploring a large corpus of job postings in recruitment market. Then, we developed a Skill Popularity based Topic Model (SPTM) for modeling the generation of skill network. A unique perspective of SPTM is that it can integrate different criteria of jobs and the hierarchical dependence of skills. Therefore, with the help of SPTM, we can effectively rank the job skills based on their multi-perspective popularity. Extensive experiments on real-world recruitment data validated the effectiveness, and also revealed some recruitment-oriented rules, which proved the potential of SPTM for measuring the popularity of job skills.

In the future, we will consider more ``personalized" factors for companies, i.e., to reveal their preference or business range of job skills, as well as latent business competition between them. Also, we plan to study more real-world applications based on the skill popularity obtained by our approach, such as resume screening and talent recommendation. Besides, more comprehensive studies will be conducted, such as the job-candidate mapping performance, and the popularity of job skills in other professional domains.

\section{ Acknowledgments}
This research was partially supported by grants from the National Natural Science Foundation of China (Grant No. U1605251, 71531001, 61703386 and 61325010), the Anhui Provincial Natural Science Foundation (Grant No. 1708085QF140), and the Fundamental Research Funds for the Central Universities (Grant No. WK2150110006).

\bibliography{sigproc}
\bibliographystyle{aaai}

\end{document}